\documentclass[11pt,twoside]{article}

\usepackage{asp2006}
\usepackage{epsf}
\usepackage{graphicx}
\usepackage{lscape}

\begin{document}
\title{SV Cen reveals its mystery}
\author{Micha{\l} Siwak$^1$, Stanis{\l}aw Zo{\l}a$^{2,3}$, Slavek Rucinski$^1$}
\affil{{\tiny $^1$Department of Astronomy and Astrophysics, University of Toronto, 50 St.George Street, M5S 3H4 Toronto\\
$^2$Astronomical Observatory of the Jagiellonian University, ul. Orla 171, 30-244 Krak{\'o}w\\
$^3$Mount Suhora Astronomical Observatory,  Pedagogical University, ul. Podchor\c{a}{\.z}ych 2, 30-084 Krak{\'o}w}}

\begin{abstract}
{\small Our very-first high resolution spectra of SV Cen close binary system obtained in the $H_{\alpha}$ line reveal its absorption and emmision components, changing with orbital phase.
An accretion disk surrounding the component eclipsed at the primary minimum is the most plausible explanation of this complex structure.}
\end{abstract}

\section*{Introduction}

The true nature of SV Cen -- one of the most famous binary system, is still not well known.
Light curve modeling programs based on the Roche model predict a contact configuration for this system, with both components almost filling the outer Roche lobe. These models predict also the largest known value of the temperature difference between components of about 9000 K, which is entirely inconsistent with the common envelope contact model.
On the other hand, the very rapid period decrease observed in the O-C diagram (\citet{kr}) suggests a semidetached rather than a contact configuration.
Led by the above inconsistencies, \citet{lin} proposed that SV Cen might be a semidetached system with the less massive component surrounded by an accretion disk.
The tittle question of their publication {\it "Does SV Centauri harbour an accretion disk?"} has prompted us to investigate this system.

\section*{Observations, data reduction and conclusions}

In a preliminary investigation we used the echelle spectrograph mounted on the 2.5 m Du Pont telescope at the Las Campanas Observatory.
In total, 12 spectra were obtained in February and March 2009.
The orbital phases were calculated using the linear ephemeris provided by \citet{kr2}.
A sample of seven heliocentric-system, rectified spectra taken in 
the $H_{\alpha}$ line is shown in Fig.1 (a-g).
This line shows strong emission components at almost all phases, 
especially close to the secondary eclipse.
To improve visibility of the emission, we decided to remove the spectrum of the secondary star.
As a template we chose the spectrum obtained at the orbital phase 0.044 (Fig.1(a)); this spectrum contains the pure spectrum of the secondary star with its characteristic rotational broadening.
Using the well orbital-phase covered radial velocity curve of SV Cen published by \citet{ruc}, we shifted this secondary spectrum by the respective RV differences to use it for division of the spectra at the orbital phases 0.35 to 0.5.
As the result, we obtained the double-peaked emission line (Fig.1 (h,i)), which is a characteristic feature of permanent accretion disks (\citet{sm}).
The disk is surrounding the primary - mass gaining component, which is visible in 
front of the secondary - mass losing star, at phases close to the secondary eclipse.

For the first time in a class of short period close binary systems, previously
 unambiguously classified as a contact binary star, 
 we see a very well defined signature of an accretion disk.
With AW UMa (\citet{prib}), this is the second case when the contact model
and the perfect fit of the EW-type light curve that it offers, fails when 
confronted with spectroscopic observations.

\begin{figure*}[]
\centerline{%
\begin{tabular}{c@{\hspace{0.0pc}}c@{\hspace{0.0pc}}c}
\includegraphics[width=1.25in, angle=-90]{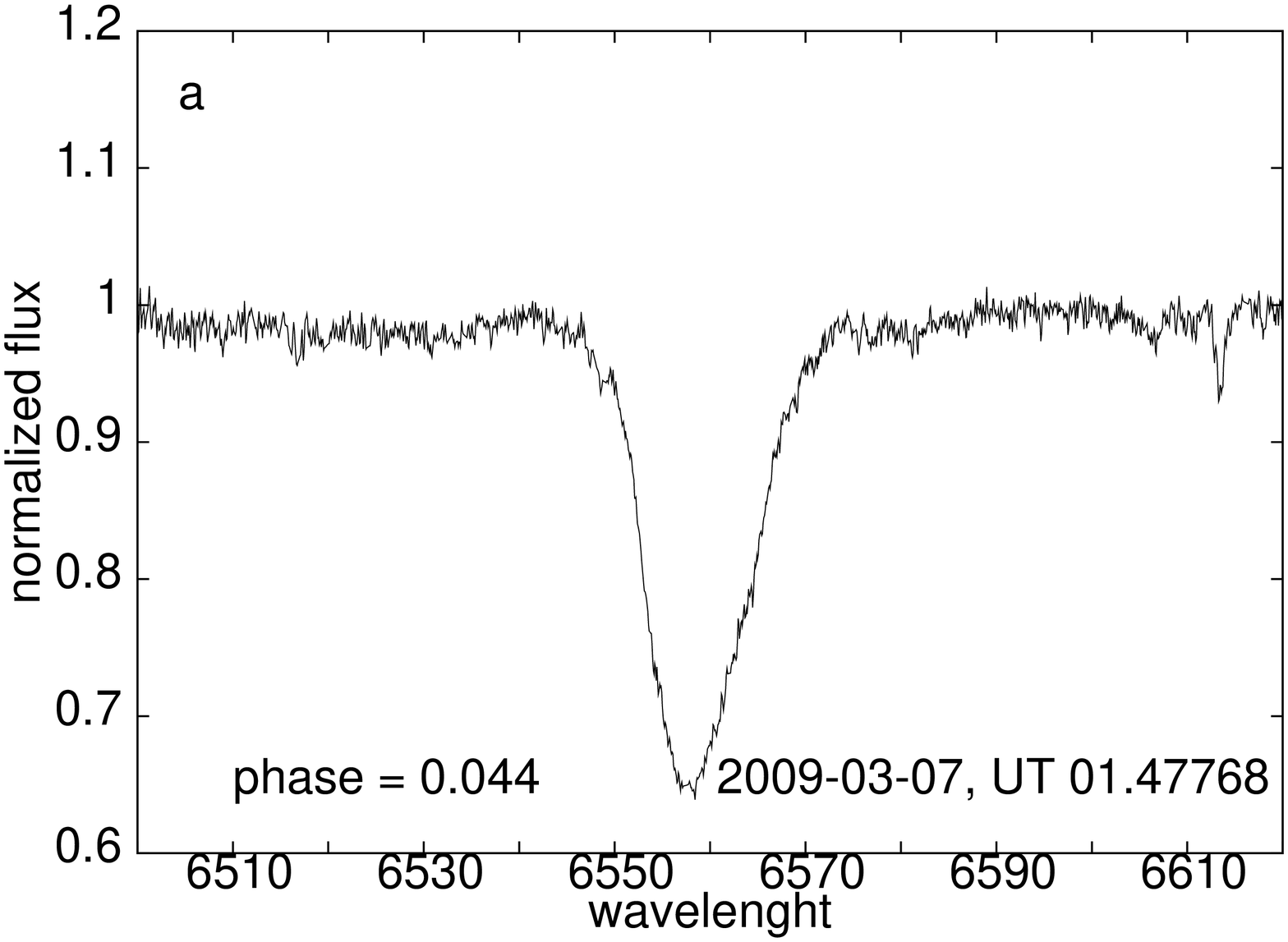} &
\includegraphics[width=1.25in, angle=-90]{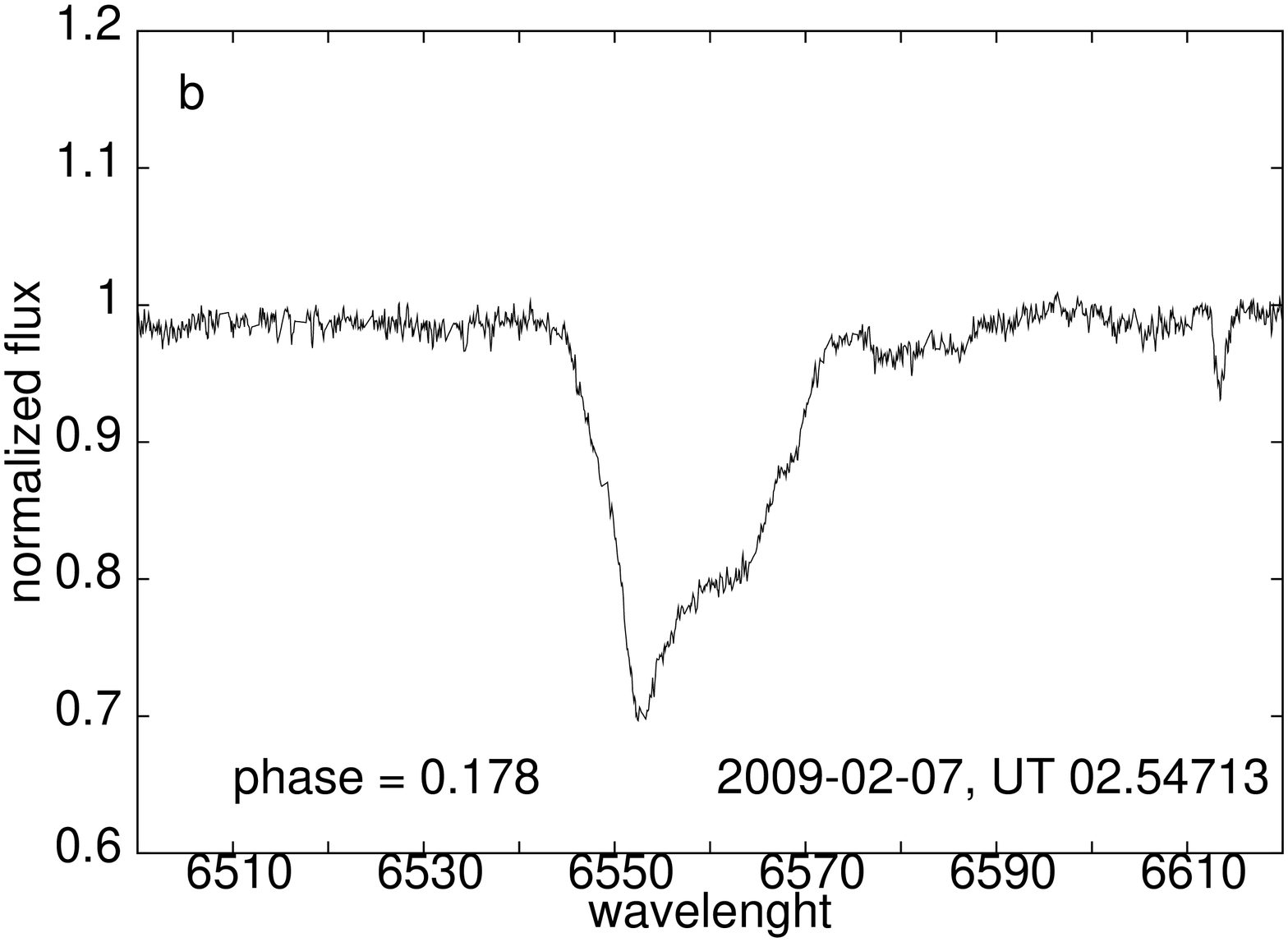} &
\includegraphics[width=1.25in, angle=-90]{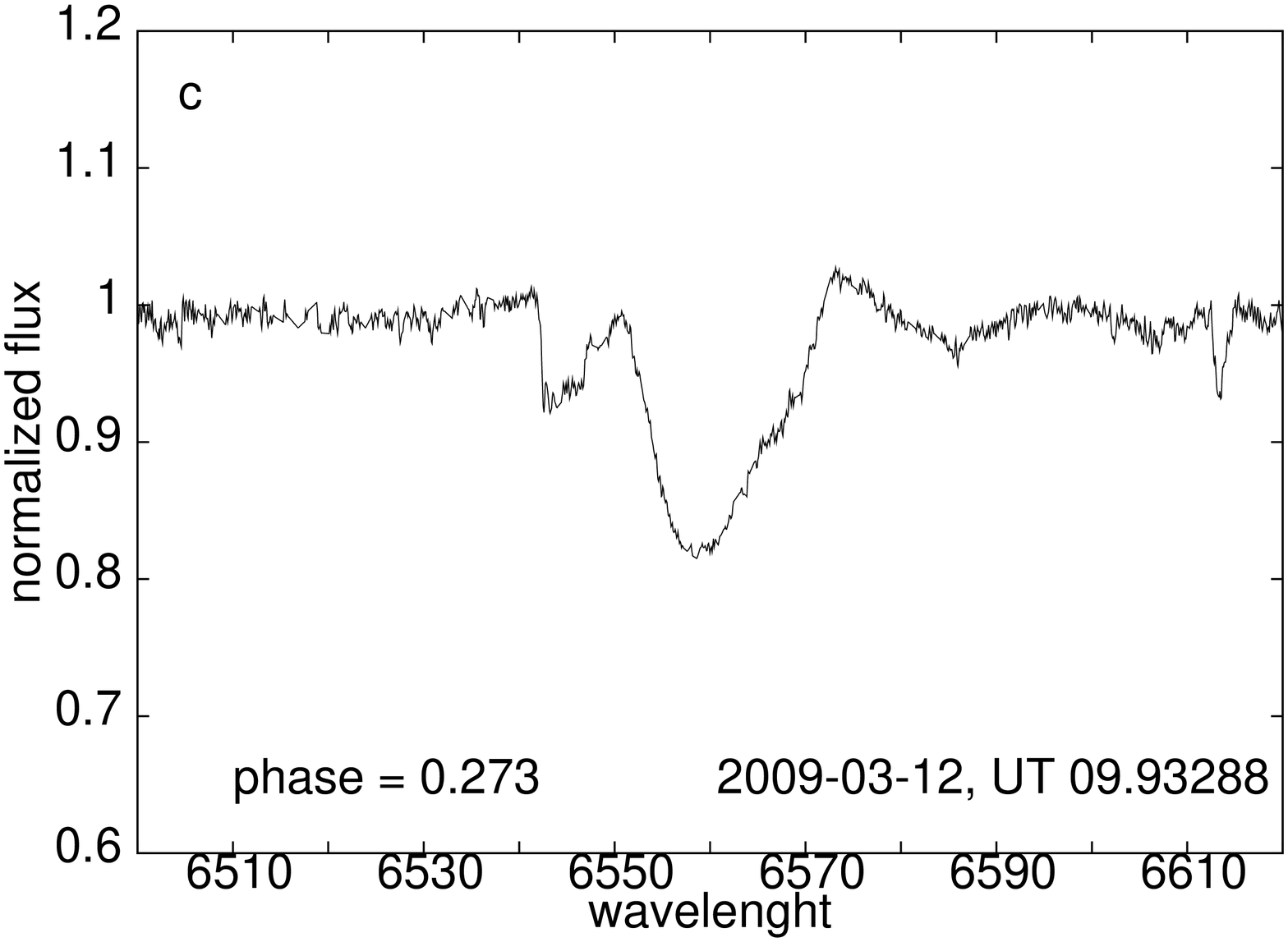} \\
\includegraphics[width=1.25in, angle=-90]{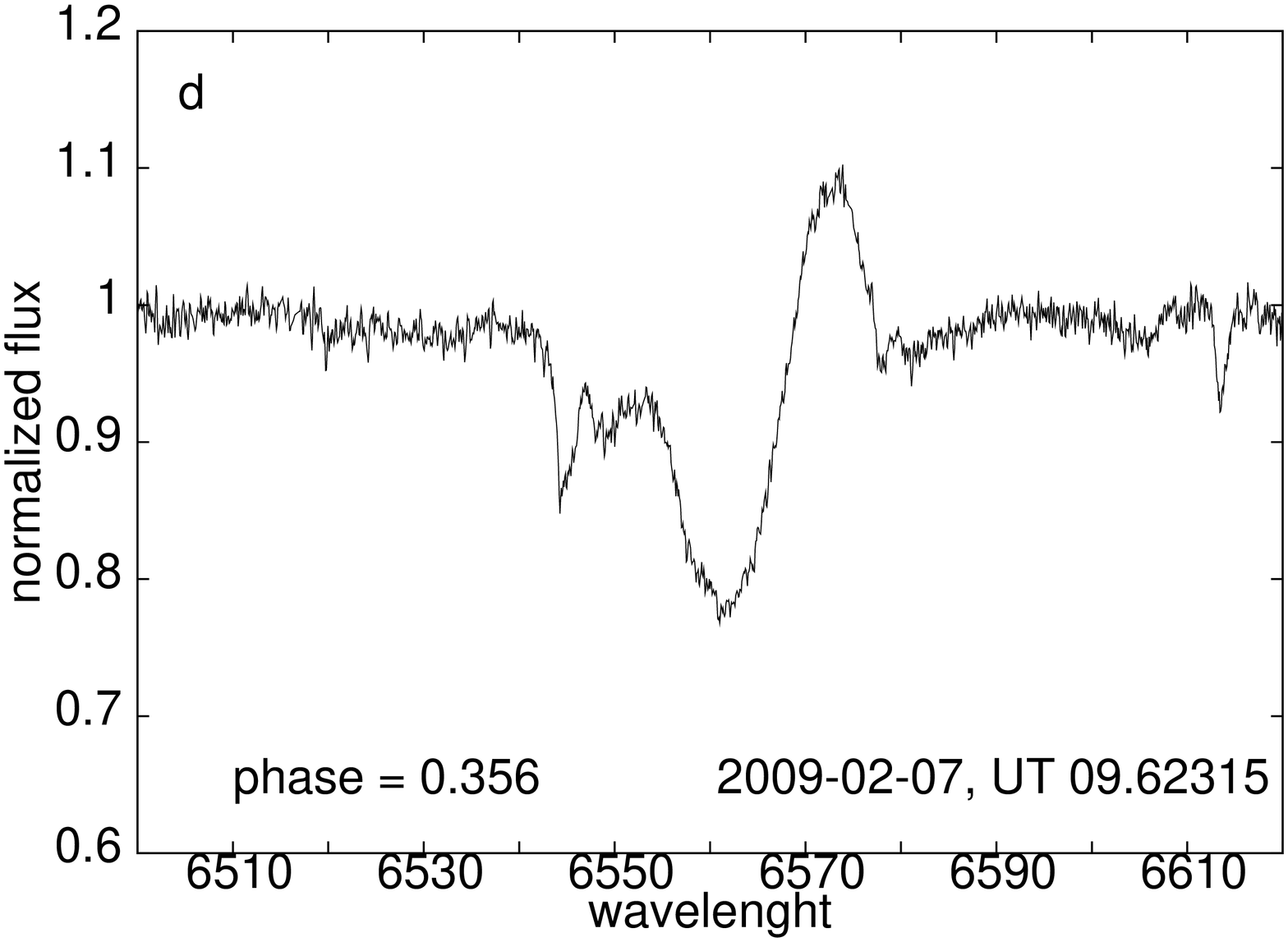} &
\includegraphics[width=1.25in, angle=-90]{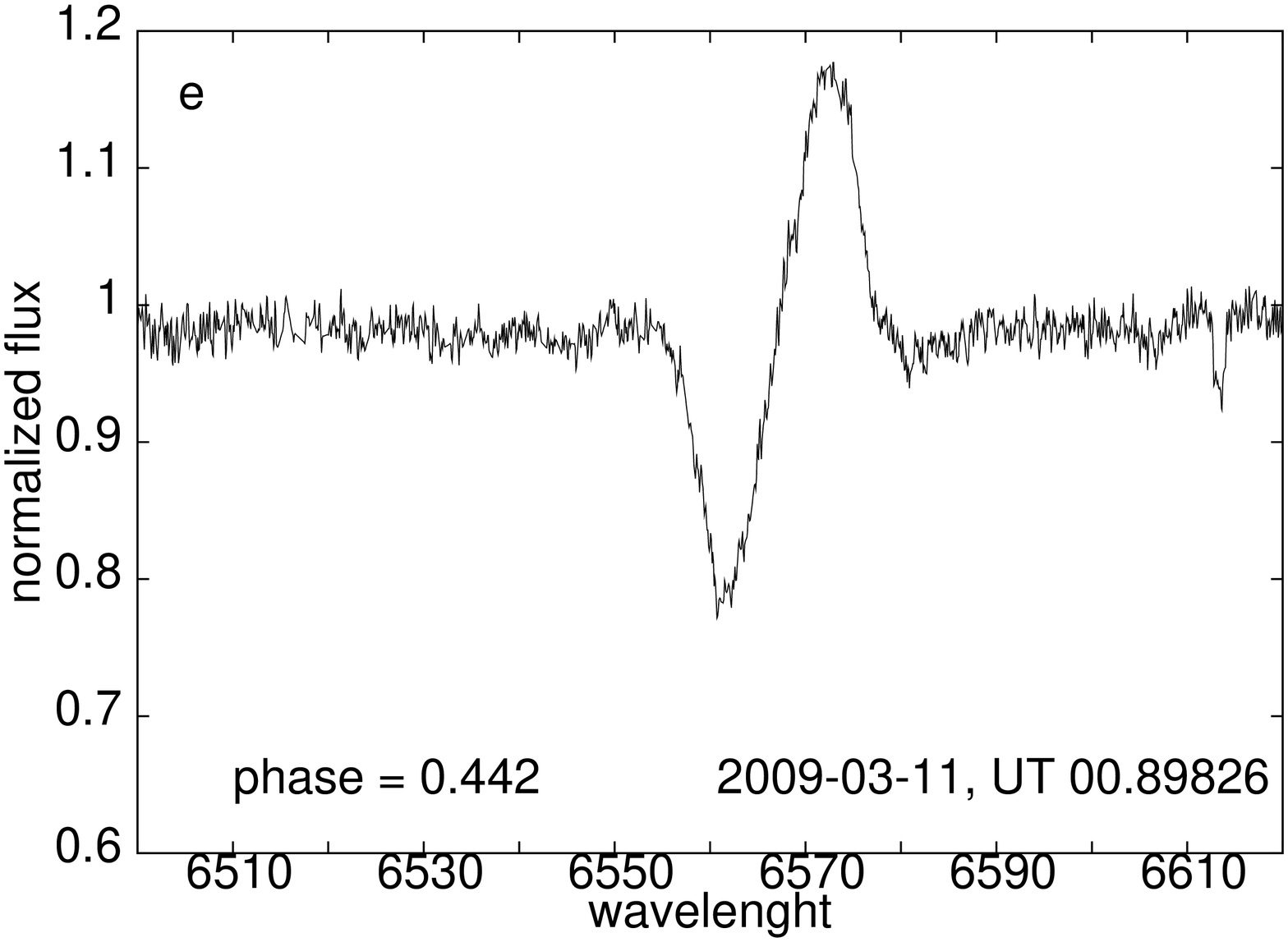} &
\includegraphics[width=1.25in, angle=-90]{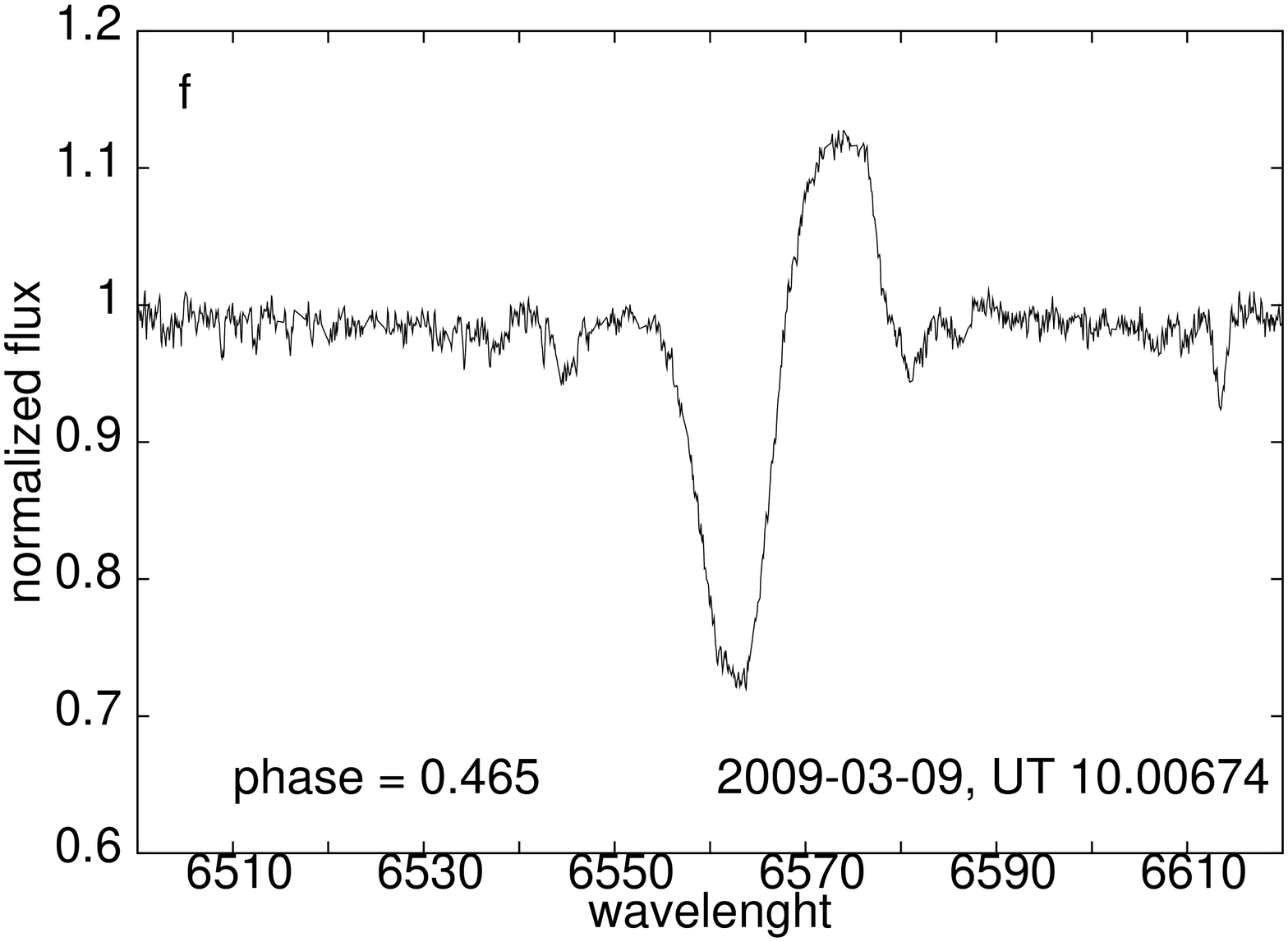} \\
\includegraphics[width=1.25in, angle=-90]{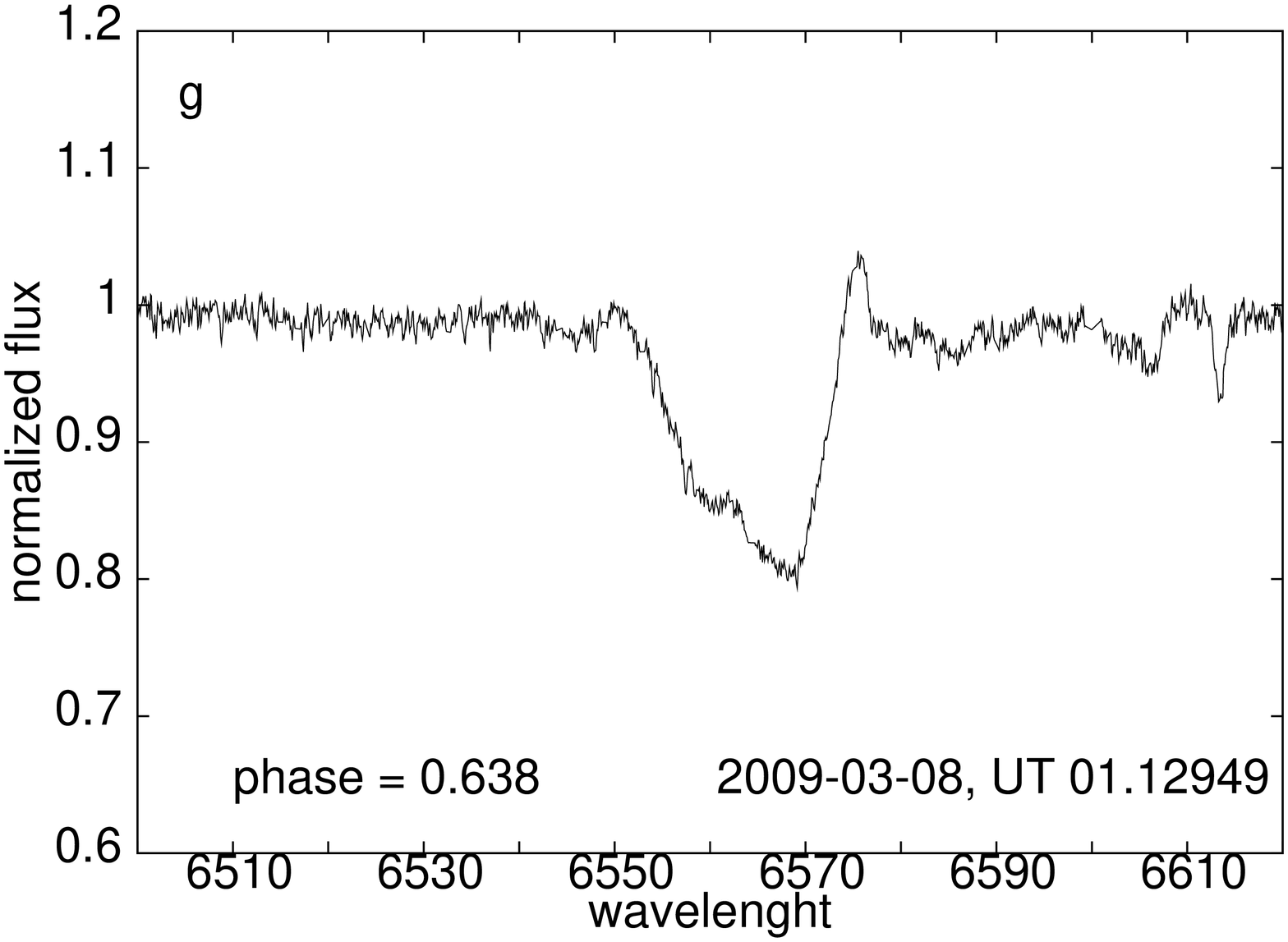} &
\includegraphics[width=1.25in, angle=-90]{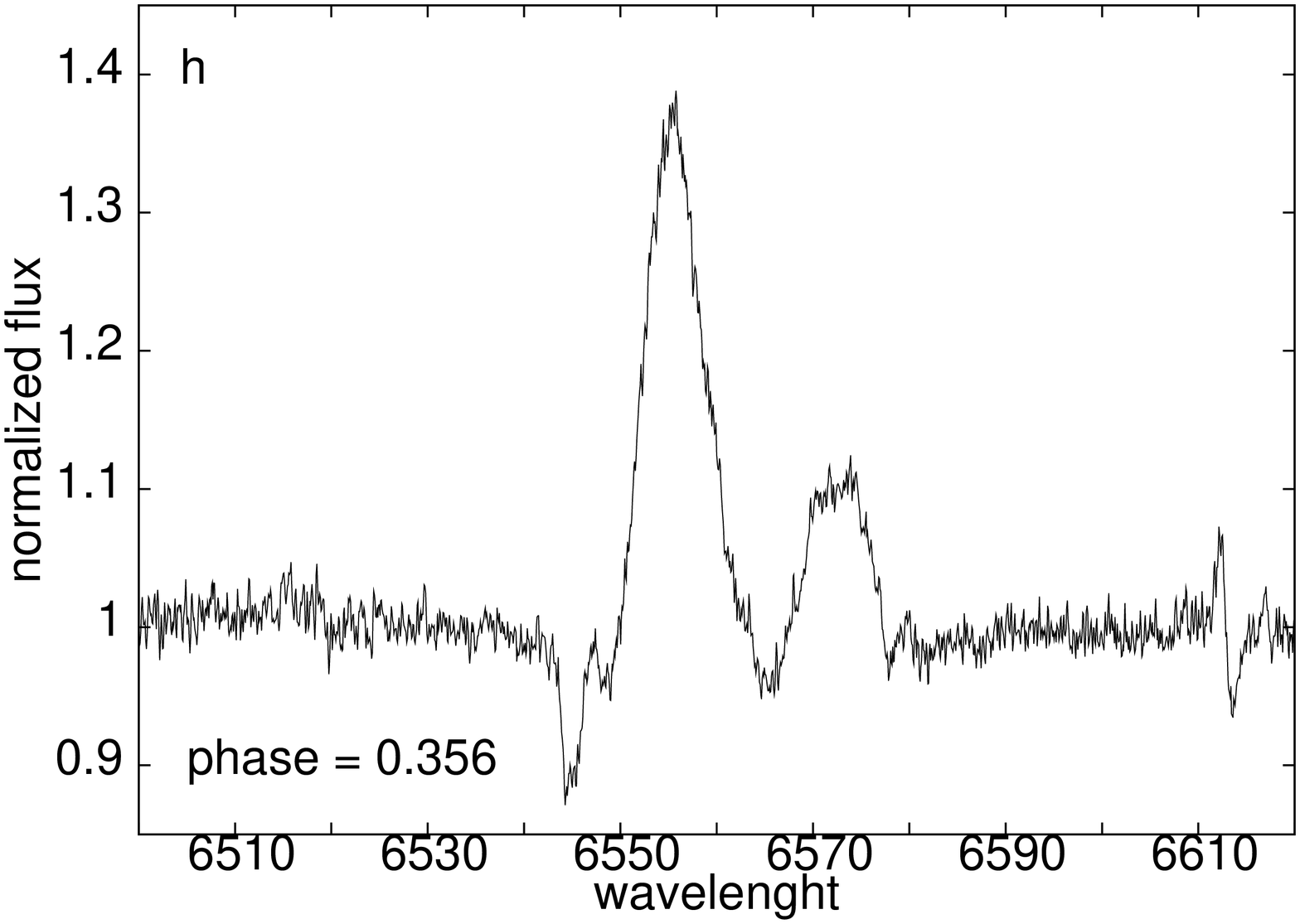} &
\includegraphics[width=1.25in, angle=-90]{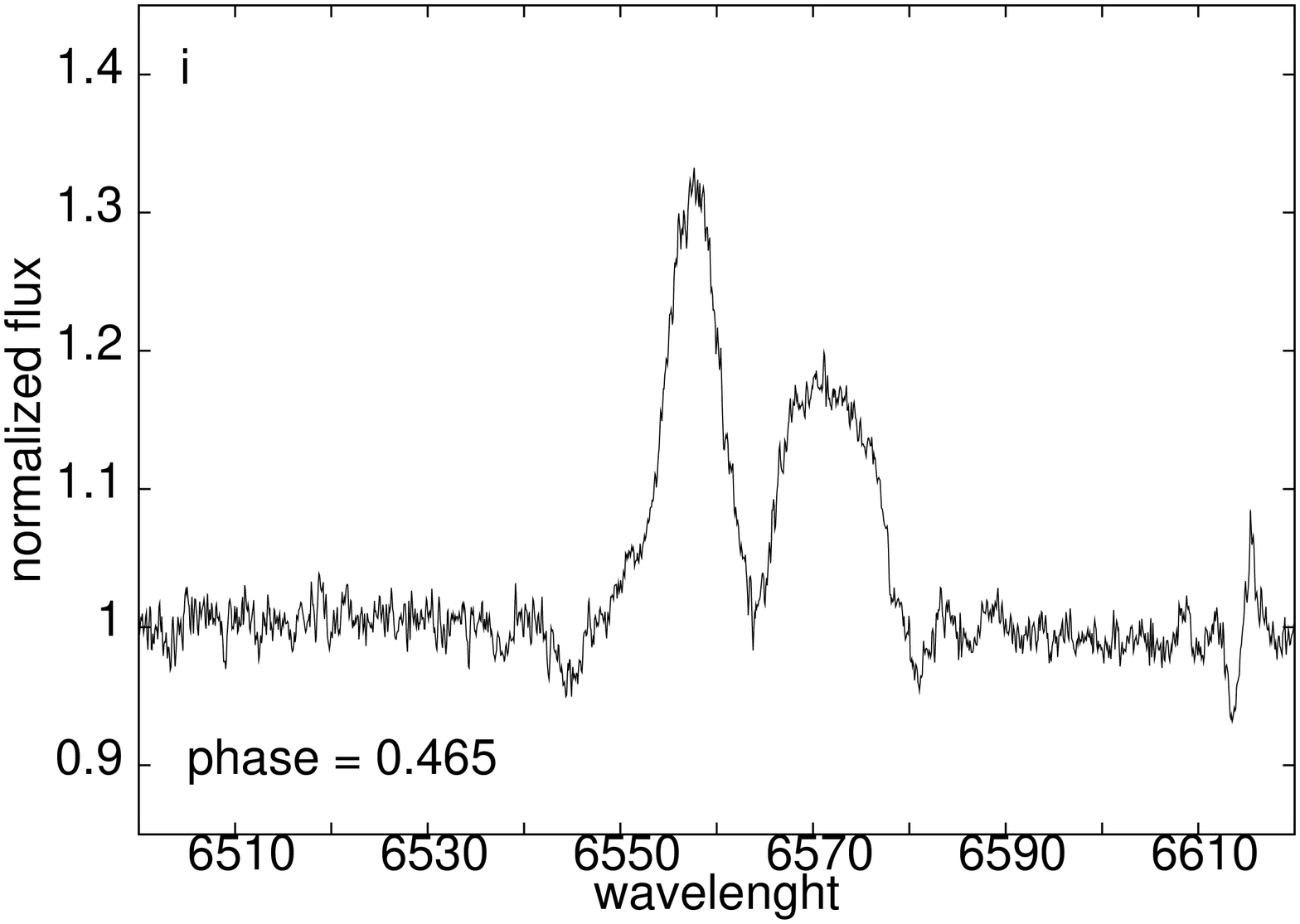}
\end{tabular}}
{\sf \caption {\small The $H_{\alpha}$ spectra of SV Cen obtained in different orbital phases. The last two panels give the contribution of the accretion disk.}}
\end{figure*}

\acknowledgements
{\small MS gratefully acknowledge the Las Campanas Observatory staff for their hospitality, especially Herman Olivares, Javier Fuentes, Oscar Duhalde and Andr{\'e}s Rivera for their night assistance and daily technical work. This study has been supported by the CSA SSEP and NSERC grants to SR.}

%%% THE BIBLIOGRAPHY
%%%
%%% CONSULT SECTION 3 OF "INSTRUCTIONS FOR AUTHORS" FOR HOW TO USE NATBIB.
%%% AUTHORS ARE ENCOURAGED TO USE EITHER THE "THEBIBLIOGRAPY" ENVIRONMENT
%%% BY UNCOMMENTING (DELETING THE "%" SYMBOL) THE COMMANDS BELOW, OR BY
%%% USING THE BIBTEX ENVIRONMENT. TO FIND OUT WHICH IS APPLICABLE TO YOUR
%%% CONTRIBUTION, CONSULT THE VOLUME EDITORS FOR YOUR PROCEEDINGS.
%%%

\end{document}